\documentclass[journal,twoside,web]{ieeecolor}
\usepackage[utf8]{inputenc}
\usepackage[T1]{fontenc}
\usepackage{amsmath}
\usepackage{amsfonts}

\usepackage{booktabs}
\usepackage{multirow}
\usepackage{makecell}
\usepackage{xcolor}
\definecolor{subsectioncolor}{RGB}{0,0,0} 

\usepackage{graphicx}
\usepackage{float}

\usepackage{microtype}
\usepackage{nicefrac}
\usepackage{soul}

\usepackage{algorithm}
\usepackage{algorithmic}

\usepackage{lipsum}

\usepackage{url}

\graphicspath{{media/}}
\begin{document}

\title{H3: A Healthcare Three-Hop Index for Physician Referral Network Prediction}

\author{Zhexi Gu\thanks{Zhexi Gu and Jiaxin Ying are with the School of Information and Library Science, University of North Carolina at Chapel Hill, Chapel Hill, NC 27599, USA (email: zhexigu@unc.edu; jiaxiny@unc.edu).}, Jiaxin Ying,
        Xu-Wen Wang\thanks{Xu-Wen Wang is the Channing Division of Network Medicine, Brigham and Women's Hospital, Harvard Medical School, Boston, MA 02115, USA (email: spxuw@channing.harvard.edu). }, and
        Can Chen\thanks{Can Chen is with the School of Data Science and Society, the Department of Mathematics, and  the Department of Biostatistics, University of North Carolina at Chapel Hill, Chapel Hill, NC 27599, USA (email: canc@unc.edu).}
       }
        

\maketitle

\begin{abstract}
Accurate prediction of physician referral links is essential for optimizing care coordination and reducing fragmentation in healthcare delivery. However, existing computational methods, ranging from triadic closure heuristics to graph neural networks, fail to capture the intrinsic properties of physician referral networks, including sparsity, disassortative degree mixing, and hub-dominated topology. Here, we propose H3, a healthcare three-hop index that addresses these limitations by modeling indirect referral pathways through intermediate physicians, with degree-based normalization and a redundancy penalty to mitigate hub-mediated noise. Using Medicare Physician Shared Patient Patterns data, we evaluate H3 under two complementary prediction regimes: within-period prediction, which assesses recovery of contemporaneous referral links under sparse conditions, and cross-period prediction, which tests robustness to temporal shift as referral windows expand. Across both regimes, H3 consistently outperforms classical heuristics and deep learning-based baselines. Unlike black-box neural network approaches, H3 produces fully decomposable predictions traceable to specific intermediary physicians, offering a transparent and deployable solution for referral network completion.
\end{abstract}


\section{Introduction}

\IEEEPARstart{H}{ealthcare} referral networks are fundamental to determining access, quality, and cost of care across the health system~\cite{an2018analysis,donker2010patient}. Within Medicare populations or systems, physician collaboration network structure significantly impacts clinical outcomes: well-connected networks correlated with reduced hospitalization costs and lower readmission rates~\cite{uddin2012effect, pollack2013patient}, while fragmented networks are associated with duplicated diagnostic testing, medication errors, and suboptimal care coordination~\cite{pham2007care}. At the physician pair level, Agha et al.~\cite{agha2022team} demonstrate that primary care providers who concentrate referrals within a smaller set of specialists build stronger team-specific capital, leading to 4\% lower healthcare utilization with no reduction in care quality. Furthermore, fragmented division of labor across organizational boundaries has been shown to increase healthcare costs, as patients whose care is spread across more organizations incur higher spending due to coordination frictions~\cite{agha2019fragmented, cebul2008organizational}. Referral path structure also carries predictive power for treatment decisions and patient outcomes in cardiovascular care, with network-derived features outperforming traditional clinical predictors~\cite{an2018referral}. These findings collectively establish that the structure of referral relationships is not merely descriptive but a critically important determinant of care quality and efficiency.

Despite the established importance of referral networks, existing research has been predominantly descriptive, like characterizing referral volume, specialty composition, and network topology, while the networks themselves are critically incomplete~\cite{xia2021effects, barnett2012physician}. Not all referral events generate observable co-patient patterns within any fixed temporal window, provider transitions continuously erode recorded linkages, and privacy-driven data suppression removes low volume dyads from public datasets~\cite{barnett2011mapping, landon2012variation}. This incompleteness is consequential because missing links distort measured topology by artificially inflating sparsity, obscuring community structure, and masking hub-mediated coordination pathways~\cite{valente2012network}. These structural gaps in turn compromise downstream analyses of care quality, cost, and fragmentation~\cite{pham2007care, agha2019fragmented}, and risks misidentifying observational gaps as true structural fragmentation. Recovering the latent structure of referral networks therefore requires link prediction~\cite{lu2011link}, which involves estimating the likelihood of unobserved relationships from the existing topology. This approach allows identification of provider dyads that are structurally predisposed to collaborate, enabling health systems to proactively design referral pathways and target care coordination interventions.

Numerous computational methods have been proposed for link prediction of general networks. Early similarity approaches, including common neighbors~\cite{liben2003link}, Jaccard coefficient~\cite{liben2003link}, and the Adamic--Adar index~\cite{adamic2003friends}, operationalize the principle of triadic closure: two nodes are likely to connect if they already share many common neighbors~\cite{holme2002growing}. Subsequent refinements such as resource allocation~\cite{zhou2009predicting} extend this intuition by weighting contributions through lower-degree intermediaries, while higher-order indices such as Katz and local path aggregate paths of increasing length. These heuristics are computationally efficient and interpretable, and they achieve strong performance in dense, assortative social networks where the triadic closure assumption holds~\cite{newman2001clustering}. However, their reliance on shared two-hop neighborhoods is a fundamental limitation in sparse, disassortative networks~\cite{arrar2024comprehensive}. In primary care provider-specialist networks with strongly negative degree mixing, two-hop indices degrade substantially~\cite{aziz2023link, wu2022link}. Although three-hop indices (e.g., L3 \cite{kovacs2019network}) outperform two-hop heuristics in low-density, disassortative regimes~\cite{zhou2021experimental}, higher-order structural methods remain challenging in healthcare applications due to the sparse and transactional nature of claims data, which limits the rich relational context these methods typically require~\cite{murala2023medmetaverse}.


The second class of methods applies graph neural network (GNN) to learn latent node representations from graph structure~\cite{kipf2016variational}. Foundational architectures such as graph convolutional network~\cite{kipf2016semi} and GraphSAGE~\cite{hamilton2017inductive} aggregate neighborhood features through spectral or spatial convolutions, enabling nodes to encode multi-hop structural context into fixed-dimensional embeddings. Subsequent models introduce attention mechanisms, as in the graph attention network~\cite{velickovic2017graph}, and more expressive aggregation schemes, as in the graph isomorphism network~\cite{xu2018powerful}, to better capture heterogeneous neighborhood distributions. For link prediction specifically, these methods are typically combined through inner products or multi-layer perceptron decoders~\cite{zhang2018link}, achieving state-of-the-art performance on benchmark citation and social graphs. More recent advances, including subgraph-based methods such as subgraph embedding for link prediction~\cite{zhang2018link} and position-aware encoding~\cite{zhang2022graph}, have further improved GNN expressiveness for link-level tasks by incorporating rich local structural features around candidate pairs rather than relying solely on standard node-level embeddings~\cite{kipf2016variational}.

Despite these advances, GNNs face three barriers that are particularly acute in the healthcare context. First, effective GNN training requires sufficiently dense and stable graph structure~\cite{zitnik2018modeling, li2022graph}. Second, clinical decision support imposes interpretability requirements that GNN latent representations cannot satisfy. Embedding-based predictions lack the explicit, auditable reasoning chains that clinicians and health system administrators need to act on predicted referral links~\cite{rudin2019stop, kawamoto2005improving}, and post-hoc explanation methods such as GNNExplainer provide only approximate, instance-level attributions rather than systematic path-based justifications. Third, scaling neighborhood aggregation to nationwide spatiotemporal graphs covering hundreds of thousands of physicians across all U.S. states and multiple temporal windows incurs substantial memory and computational cost, limiting practical deployment at scale~\cite{hamilton2017inductive}.

\begin{figure*}[t]
\centering
\includegraphics[width=0.9\textwidth]{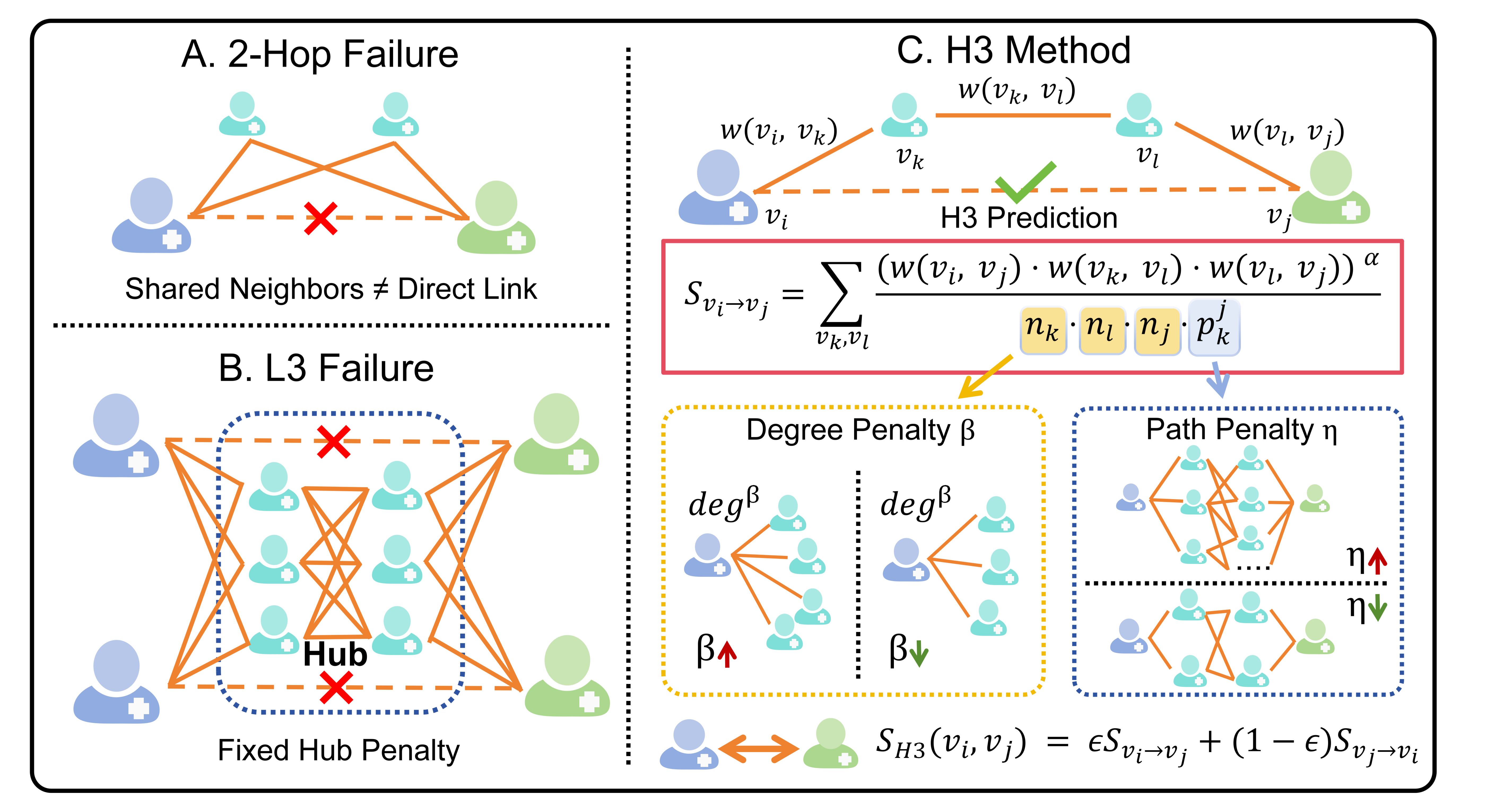}
\caption{Illustration of the H3 scoring framework.
(A)~Two-hop methods fail when $v_i$ and $v_j$ share no common neighbor, missing clinically meaningful indirect referral links.
(B)~L3 extends to three-hop paths but applies a fixed hub penalty ($\beta = 0.5$), insufficient for the heavy-tailed degree distribution of physician referral networks.
(C)~H3 scores each node pair by aggregating weighted three-hop paths $v_i \!\to\! v_k \!\to\! v_l \!\to\! v_j$. The denominator combines a tunable degree penalty $\beta$ on intermediate hubs ($n_k, n_l$), a target normalization term $n_j$, and a path redundancy penalty $p_k^j$ controlled by $\eta$ that discounts parallel paths through the same intermediary. The final undirected score $S_{\text{H3}}(v_i, v_j)$ symmetrizes the asymmetric directed scores via a convex combination controlled by $\epsilon$.}
\label{fig:h3_illustration}
\end{figure*}

To address these challenges, we introduce H3, a healthcare three-hop index designed specifically for referral networks.  H3 overcomes the identified limitations in three ways. First, it moves beyond two-hop similarity by aggregating degree-normalized three-hop paths, capturing the indirect referral sequences through intermediaries that two-hop heuristics structurally cannot detect. Second, it incorporates a redundancy penalty that suppresses spurious multi-hop paths arising from incidental patient mobility through hub health systems, directly addressing the weight heterogeneity and structural noise specific to claims-derived data. Third, each predicted link is traceable to specific intermediary physicians and patient flow volumes, providing the explicit, auditable path-based evidence that clinical decision support requires and that GNN-based latent representations cannot offer. 
We evaluate H3 on the CMS Physician Shared Patient Patterns (PSPP) dataset~\cite{an2018analysis, cms_pspp2015}. We test H3 across all U.S. states and multiple temporal windows (30-day, 90-day, and 180-day intervals), comparing it with traditional heuristics and graph-based deep learning baselines under both within-period and cross-period prediction regimes. H3 consistently outperforms existing methods, while remaining inherently interpretable as a structural heuristic, with robustness rooted in an inductive bias aligned with the sparse, disassortative, and hub-mediated structure of healthcare networks. By accurately recovering latent referral pathways, H3 has the potential to effectively inform targeted care coordination, reduce fragmentation, and improve health system efficiency.

The remainder of this article is organized as follows. Section \ref{sec:prelim} presents preliminaries, including the network representation of physician referral networks and link prediction. Section \ref{sec:meth} introduces H3, detailing its scoring algorithm, computational optimizations, and advantages over existing methods. Section \ref{sec:num} evaluates H3 across all U.S. states and multiple temporal windows  using the PSPP dataset. Section \ref{sec:dis} presents a case study and discusses the generalizability and limitations of H3, with Section \ref{sec:conclusion} concluding the work.

\section{Preliminaries}\label{sec:prelim}

\subsection{Network Representation}

A network is formally defined as $\mathcal{G} = \{\mathcal{V}, \mathcal{E}, w\}$, where 
$\mathcal{V} = \{v_1, v_2, \ldots, v_n\}$ is the  set of nodes, 
$\mathcal{E} \subseteq V \times V$ is a set of link encoding pairwise relationships between nodes, 
and $w : \mathcal{V} \times \mathcal{V} \to \mathbb{R}_{\geq 0}$ is a weight function assigning a non-negative strength to each link. The weight function allows the network to capture not only the presence of a relationship but also its magnitude or intensity.  A network can be represented by weighted adjacency matrix 
$\textbf{W} \in \mathbb{R}_{\geq 0}^{n \times n}$ such that
\begin{equation*}
\mathbf{W}_{ij} =
\begin{cases}
w(v_i, v_j), & (v_i, v_j) \in \mathcal{E} \\
0, & \text{otherwise}
\end{cases}.
\end{equation*}
The degree of a node $v_i$ are defined as
\begin{equation*}
    \deg_w(v_i) = \sum_{j=1}^n \textbf{W}_{ij}.
\end{equation*}
In an unweighted network, the weight function reduces to a binary indicator that simply records whether an link exists between two nodes. The corresponding binary adjacency matrix is $\textbf{A} = \textbf{1}[(v_i, v_j) \in \mathcal{E}]\in\mathbb{R}^{n\times n}$, and the unweighted degree of node $v_i$ is defined as $d_i = \sum_{j} \textbf{A}_{ij}$.







In physician referral networks, nodes represent individual healthcare providers, links capture co-patient relationships observed within a defined temporal window, and link weights $w(v_i, v_j)$ quantify the number of patients shared between each provider pair. Larger weights indicate stronger collaboration and more frequent coordination of care, allowing the network to reflect not only the existence of referral relationships but also their intensity and relative importance within the healthcare system. However, these networks are inherently incomplete. Not all referral events generate observable co-patient patterns within any fixed temporal window, provider transitions continuously erode recorded linkages, and privacy-driven data suppression removes low-volume dyads from public datasets.

\subsection{Link Prediction}

Given an observed network $\mathcal{G}= \{\mathcal{V}, \mathcal{E}, w\}$, the link prediction task is to estimate the likelihood of a link between each pair of nodes $(v_i, v_j) \notin \mathcal{E}$~\cite{lu2011link}. Formally, a link prediction method defines a scoring function $S : \mathcal{V} \times \mathcal{V} \to \mathbb{R}$, where a higher score $S(v_i, v_j)$ indicates a greater probability that a link either exists latently or will form between $v_i$ and $v_j$. Node pairs are then ranked according to their scores, with the top-ranked pairs treated as predicted links. 

Classical structural similarity indices derive these scores purely from the topology of $\mathcal{G}$, without requiring node attributes or external features. Two-hop indices, such as common neighbors, quantify the overlap of immediate neighborhoods:
\begin{equation*}
S_{\text{CN}}(v_i, v_j) = |\Gamma(v_i) \cap \Gamma(v_j)|,
\end{equation*}
where $\Gamma(\cdot)$ denotes the set of neighbors of a node and $|\cdot|$ denotes the set cardinality. Refinements such as Adamic--Adar~\cite{adamic2003friends} and resource allocation~\cite{zhou2009predicting} downweight contributions from high-degree shared neighbors, using $1/\log(d_k)$ or $1/d_k$ for each intermediary $v_k \in \Gamma(v_i) \cap \Gamma(v_j)$. These approaches rely on the principle of triadic closure, assuming that nodes sharing neighbors are likely to form a link. However, this assumption often fails in sparse, disassortative networks where immediate neighbors are limited and most meaningful interactions occur via longer paths. This is particularly true in biological networks and healthcare referral networks, where informative connections frequently span three hops rather than two.

To address this, Kovács et al.\cite{kovacs2019network} proposed the L3 index, a three-hop structural similarity measure originally designed for protein interaction networks. L3 evaluates a candidate pair $(v_i,v_j)$ by aggregating over all three-hop paths $v_i \to v_k \to v_l \to v_j$, normalizing each path by the unweighted degrees of its two intermediate nodes:
\begin{equation*}
S_{\text{L3}}(v_i, v_j) =
\sum_{v_k \in \Gamma(v_i)}
\sum_{v_l \in \Gamma(v_j)}
\frac{\textbf{A}_{kl}}{\sqrt{d_k d_{l}}},
\end{equation*}
where $d_{k}$ and $d_{v_l}$ are the unweighted degrees of nodes $v_k$ and $v_l$, and $\textbf{A}$ is the binary adjacency matrix. L3 has been shown to outperform two-hop indices on networks with bipartite-like, disassortative structure, providing a theoretical justification based on degree heterogeneity and low local clustering. However, directly applying L3 to physician referral network prediction presents several challenges, including sparsity, disassortative degree mixing, and a hub-dominated topology, which motivate the development of H3.

\section{Method}\label{sec:meth}

In this section, we introduce H3, a healthcare three-hop index designed specifically for physician referral networks. H3 builds on the intuition of three-hop structural similarity while addressing the unique challenges of healthcare networks, including extreme sparsity, hub-dominated degree distributions, and weighted, clinically meaningful links. It aggregates multi-step referral pathways, incorporates normalization to account for node degree and patient-sharing volume, and provides fully interpretable, path-level scores that link each predicted connection to specific intermediary providers. This combination of structural rigor, weight sensitivity, and interpretability makes H3 particularly suited for accurate and clinically actionable link prediction in physician referral networks. Detailed workflow of H3 and its distinctions from 2-hop and L3 methods are presented in Fig. \ref{fig:h3_illustration}.

\subsection{H3 Scoring Algorithm}
Let $\mathcal{G} = \{\mathcal{V}, \mathcal{E}, w\}$ be an undirected, weighted network derived from patient-sharing records between physicians, where $w(v_i,v_j)$  denotes the number of patients shared between physician sharing between physician $v_i$ and physician $v_j$. Capturing multi-step referral patterns is crucial because clinically meaningful connections often occur through indirect pathways, especially in sparse networks where two physicians may not share patients directly but are connected via intermediaries. Inspired by multi-hop link prediction strategies~\cite{sun2025review, zhang2018link, kovacs2019network}, we define a three-hop scoring function H3 to quantify the referral affinity between node $v_i$ and $v_j$ ($i\neq j$). Although the network is undirected, we enumerate three-hop paths in a fixed directional order $v_i \to v_k \to v_l \to v_j$ to define an initially asymmetric score:
\begin{equation*}
\scalebox{0.88}{$\displaystyle
    S_{v_i \to v_j} = 
\sum_{v_k \in \Gamma(v_i)} \;\sum_{v_l \in \Gamma(v_k) \cap \Gamma(v_j)}
\frac{\Big(w(v_i,v_k)\, w(v_k,v_l)\, w(v_l,v_j)\Big)^{\alpha}}
{n_{k}\, n_{l}\, n_{j}\, p_{k}^{(j)}}$}.
\end{equation*}
The summation is restricted to intermediate nodes $v_k \in \Gamma(v_i)$ and connectors $v_l \in \Gamma(v_k) \cap \Gamma(v_j)$ ensuring that only valid three-hop paths contribute to the score. Each component of the denominator serves a structural purpose:
\begin{align*}
    n_{k} &= \max\bigl(\deg_w(v_k), 1\bigr)^{\beta}, \\
n_{l} &= \max\bigl(\deg_w(v_l), 1\bigr)^{\beta}, \\
n_{j} &= \max\bigl(\deg_w(v_j), 1\bigr)^{\gamma}, \\
p_{k}^{(j)} &= \max\bigl((\log(1 + q_{k}^{(j)}))^{\eta}, p_{\min}\bigr),
\end{align*}
where $q_{k}^{(j)}$ counts the number of distinct connectors $v_l$ forming valid three-hop paths from $v_i$ to $v_j$, the exponents $\alpha, \beta, \gamma, \eta\in(0,1]$  control the relative influence of path strength, hub suppression, target normalization, and redundancy penalization, and   $p_{\min}\in(0,1]$ is  the floor constant.

The final H3 score is symmetrized to account for bidirectional referral potential:
\begin{equation}\label{eq:h3_directed}
    S_\text{H3}(v_i, v_j) = \epsilon S_{v_i \to v_j} + (1 - \epsilon ) S_{v_j \to v_i}
\end{equation}
for $\epsilon\in [0,1]$. This formulation addresses several challenges inherent to physician referral networks. Aggregating three-hop paths identifies indirect referral pathways missed by two-hop methods, capturing clinically meaningful connections in sparse networks. Unlike protein interaction networks where node degrees follow an approximately Poisson distribution, physician referral networks exhibit a heavy-tailed degree distribution in which a small number of high-volume hub providers concentrate a disproportionate share of patient flow. The tunable exponent $\beta$ allows the hub suppression to be calibrated to this skew rather than applying the fixed $\beta = 0.5$ normalization of L3. Similarly, because hub providers connect to many specialists across unrelated care pathways, multiple three-hop paths through the same intermediary $v_k$ often reflect structural co-location rather than genuine referral affinity. The redundancy penalty $p_k^{(j)}$ discounts such parallel paths and prevents intermediary similarity from inflating the final score. Raising link weights to $\alpha$ preserves the relative strength of patient-sharing relationships, emphasizing more significant collaborations. Collectively, these design choices allow H3 to capture robust, interpretable, and clinically relevant referral patterns beyond the scope of traditional two-hop or unweighted three-hop measures such as L3.

\subsection{Computational Optimization}

The naive computation of H3 using \eqref{eq:h3_directed} requires enumerating all three-hop paths using triple-nested loops, which is computationally prohibitive for large-scale physician networks. To address this, we present a set of optimizations that enable efficient scoring on networks with millions of links.

\subsubsection{Sparse Matrix Acceleration}

To achieve scalability, we reformulate the scalar summation in \eqref{eq:h3_directed} as a sequence of sparse matrix operations. Let $\textbf{W} \in \mathbb{R}^{n \times n}$ denote the weighted adjacency matrix of the physician network. We first define the element-wise powered matrix $\tilde{\textbf{W}} = \textbf{W}^{\circ \alpha}$ where 
 $(\cdot)^{\circ \alpha}$ denotes the Hadamard (element-wise) power operation. Because the numerator of \eqref{eq:h3_directed} raises the product of three link weights to $\alpha$, and $(abc)^{\alpha} = a^{\alpha}\, b^{\alpha}\, c^{\alpha}$ for non-negative reals, applying the exponent element-wise to each edge weight before matrix multiplication yields an identical result
\begin{equation*}
\Big(w(v_i,v_k)\, w(v_k,v_l)\, w(v_l,v_j)\Big)^{\alpha}
= \tilde{\textbf{W}}_{ik}\, \tilde{\textbf{W}}_{kl}\, \tilde{\textbf{W}}_{kj}.
\end{equation*}
This equivalence enables the path-weight aggregation to be computed via successive sparse matrix multiplications over $\tilde{\textbf{W}}$.

Next, we construct the diagonal degree matrix $\textbf{D}_{\text{conn}} \in \mathbb{R}^{n \times n}$, with entries
\begin{equation*}
[\textbf{D}_{\text{conn}}]_{ii} = \max\bigl(\deg_w(v_i), 1\bigr)^{\beta}.
\end{equation*}
We then define left- and right-normalized transition matrices:
\begin{equation*}
\textbf{W}_{\text{L}} = \tilde{\textbf{W}} \, \textbf{D}_{\text{conn}}^{-1}, \quad
\textbf{W}_{\text{R}} = \textbf{D}_{\text{conn}}^{-1} \, \tilde{\textbf{W}}.
\end{equation*}
The unpenalized score matrix, which aggregates all three-hop path contributions without the $p_k^{(j)}$ penalty, can be computed via sparse matrix multiplication
\begin{equation*}
\textbf{S}_{\text{raw}} = \textbf{W}_{\text{L}}  \textbf{W}_{\text{R}}  \tilde{\textbf{W}}  \textbf{D}_{\text{target}}^{-1},
\label{eq:matrix_form}
\end{equation*}
where $\textbf{D}_{\text{target}} \in \mathbb{R}^{n \times n}$ is the diagonal matrix with entries
\begin{equation*}
[\textbf{D}_{\text{target}}]_{jj} = \max\bigl(\deg_w(v_j),\, 1\bigr)^{\gamma},
\end{equation*}
corresponding to the target normalization factor $n_j$. Right-multiplying by $\mathbf{D}_{\text{target}}^{-1}$ scales each column $j$ of the preceding product, which is equivalent to dividing every path score terminating at target node  by $n_j$, consistent with the per-path normalization. The successive multiplications  correspond to the hops $v_i \to v_k$, $v_k \to v_l$, and $v_l \to v_j$, with degree normalization absorbed into the transition matrices. This formulation eliminates the explicit triple-nested loop over intermediate nodes, leveraging optimized sparse BLAS kernels for efficient computation.

\subsubsection{Structural Precomputation and Caching}

The path multiplicity penalty $p_k^{(j)}$ introduces a data-dependent term that cannot be directly absorbed into the matrix formulation. Specifically, computing $q_k^{(j)}$, the number of distinct connectors $v_l$ such that $(v_k, v_l) \in \mathcal{E}$ and $(v_l, v_j) \in \mathcal{E}$, requires knowledge of the two-hop neighborhood structure. We observe that $q_k^{(j)}$ corresponds to the $(k, j)$th entry of the squared binary adjacency matrix, i.e., 
$
q_k^{(j)} = [\textbf{A}^2]_{kj}
$
where $\textbf{A} \in \{0,1\}^{n \times n}$ is the unweighted adjacency matrix. Rather than computing the dense $\textbf{A}^2$, we exploit sparsity by precomputing only the non-zero entries corresponding to valid two-hop pairs in the training network. This selective computation reduces both memory footprint and runtime, as the number of such pairs scales with $\mathcal{O}(\texttt{nnz}(\textbf{A}^2))$ rather than $\mathcal{O}(n^2)$, where \texttt{nnz} denotes the MATLAB operator that counts nonzero entries in a matrix. The penalty matrix $\mathbf{P} \in \mathbb{R}^{n \times n}$ is then constructed with entries
\begin{equation*}
[\textbf{P}]_{kj} = \max\Bigl(\bigl(\log(1 + [\textbf{A}^2]_{kj})\bigr)^{\eta},\, p_{\min}\Bigr)=p_k^{(j)},
\end{equation*}
and the final score matrix is obtained via element-wise division
\begin{equation*}
\textbf{S} = \textbf{S}_{\text{raw}} \oslash \mathbf{P},
\end{equation*}
where $\oslash$ denotes the Hadamard division, applied only over the sparse support of $\textbf{S}_{\text{raw}}$. This approach efficiently incorporates the redundancy penalty while preserving the sparsity and scalability of the computation.

\subsubsection{Complexity Analysis}

Table~\ref{tab:complexity} summarizes the computational complexity of each algorithmic component, comparing the naive triple-loop implementation against our optimized sparse matrix formulation. Here, $n$ denotes the number of nodes, $|E|$ the number of link, $d$ the average node degree, and $L$ the number of non-zero entries in intermediate products. In practice, the optimized implementation achieves significant speedup on physician networks with $n > 10^4$ nodes, reducing wall-clock time from hours to minutes.

\begin{table}[t]
\centering
\caption{Complexity comparison between naive and optimized implementations. SpMM stands for sparse matrix–matrix multiplication.}
\label{tab:complexity}
\begin{tabular}{lccc}
\toprule
\textbf{Step} & \textbf{Naive} & \textbf{Optimized} & \textbf{Space} \\
\midrule
Path Enumeration & $O(n  d^3)$ & $O(L  d^2)$ & $O(|E|)$ \\
Penalty Calculation & $O(n  d^2)$ & $O(\texttt{nnz}(\textbf{A}^2))$ & $O(|E|_{\text{2-hop}})$ \\
Score Aggregation & $O(n^2)$ & $O(\text{SpMM})$ & $O(|E|_{\text{3-hop}})$ \\
\bottomrule
\end{tabular}
\end{table}

\subsection{Advantages of H3 over Existing Methods}

H3 builds on the three-hop indexing framework introduced by L3~\cite{kovacs2019network}, incorporating targeted modifications to address the structural and operational characteristics of physician referral networks. First, to mitigate the impact of redundant multi-hop paths, H3 introduces a redundancy penalty $p_k^{(j)}$ that discounts each intermediate node $v_k$ in proportion to the number of distinct connectors $v_l$ through which it can reach the target $v_j$. This mechanism addresses a common artifact in claims-derived data, where large academic medical centers and multi-specialty practices generate high volumes of incidental co-occurrence paths. By penalizing structurally redundant paths rather than uniformly attenuating high-degree nodes, H3 preserves signal from clinically meaningful referral chains while suppressing noise arising from system-level patient routing.

Second, H3 improves path aggregation by incorporating link weights raised to a power. Unlike L3, which treats all three-hop paths equally, this formulation ensures that paths supported by higher patient-sharing volumes contribute proportionally more to the overall score. In datasets such as the CMS PSPP network, where link weights span several orders of magnitude, this distinction is critical: three-hop paths through intermediaries sharing tens of patients with both endpoints provide much stronger evidence of referral affinity than paths supported by a single shared patient, a nuance that L3 cannot capture.

Finally, H3 enhances interpretability by making each predicted link $(v_i, v_j)$ fully decomposable into its constituent three-hop paths, annotated with the corresponding patient-sharing volumes $w(v_i, v_k)$, $w(v_k, v_l)$, and $w(v_l, v_j)$. This decomposition provides explicit, auditable reasoning, which is essential for clinical decision support~\cite{rudin2019stop, kawamoto2005improving}. Health system administrators can identify not only which provider pairs are predicted to collaborate but also the specific intermediary physicians and patient flow volumes that underpin each prediction, enabling targeted and evidence-based care coordination interventions. Collectively, these design enhancements allow H3 to capture robust, interpretable, and clinically meaningful referral patterns that are not adequately represented by traditional two-hop heuristics or unweighted three-hop measures.

\section{Results}\label{sec:num}
We evaluate H3 on the CMS Physician Shared-Patient Patterns (PSPP) dataset~\cite{an2018analysis, cms_pspp2015}, covering all U.S. states for the years 2014 and 2015. The PSPP dataset is derived from Medicare claims and captures de-identified patient-sharing relationships between physicians, providing a large-scale and detailed view of referral patterns across specialties and regions. The associated code can be found in \url{https://github.com/ZachGu-00/H3}.

\begin{figure*}[t]
\centering
\includegraphics[width=0.9\textwidth]{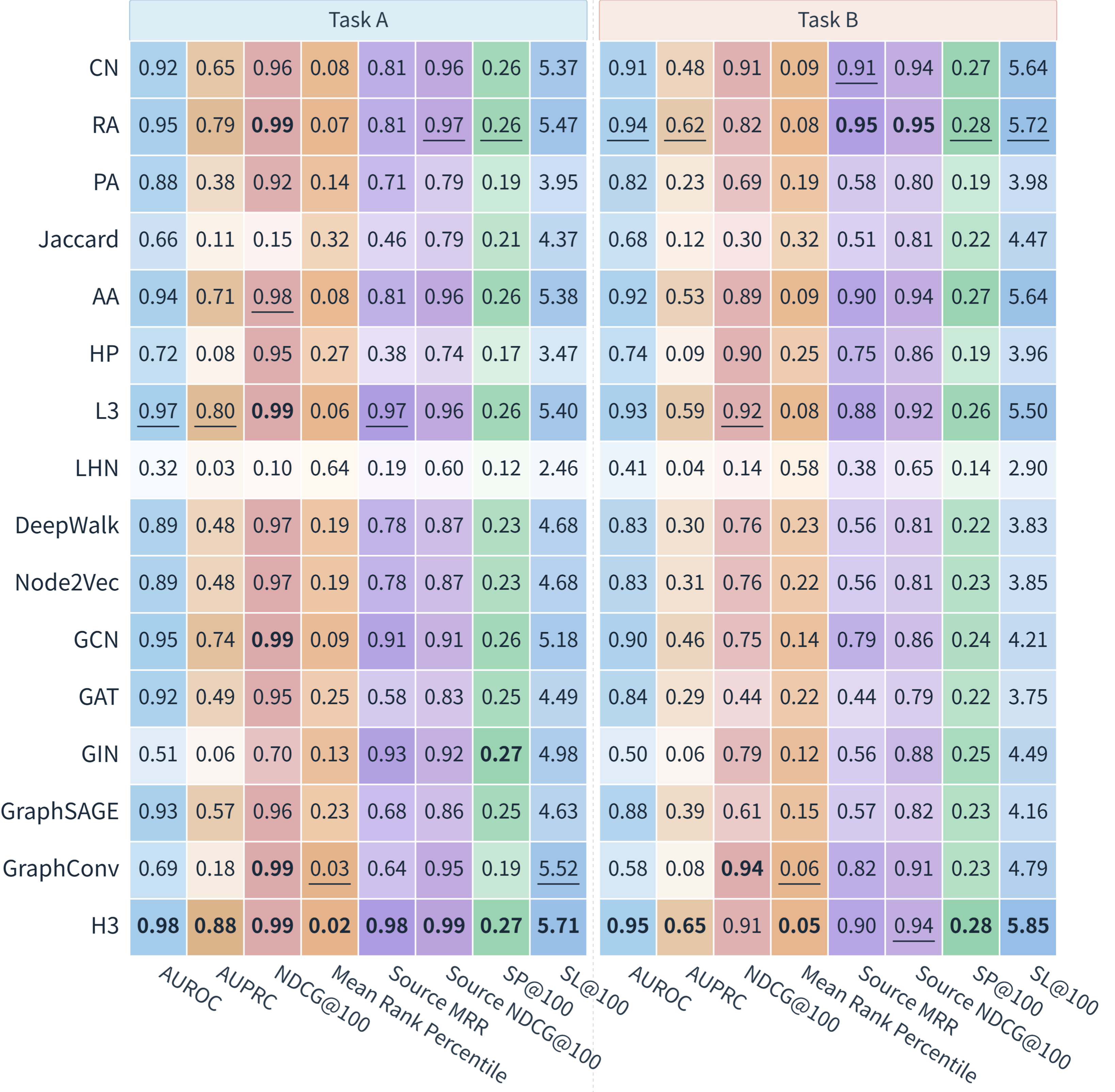}
\caption{Link prediction performance across Task~A (within-period) and Task~B (cross-period). Best results are bold; second-best are underlined.}
\label{fig:link_prediction_heatmap}
\end{figure*}
\subsection{Experiment Setup}

For each state and temporal window, we construct physician co-occurrence networks where nodes represent individual physicians and weighted links reflect shared-patient relationships observed within the specified time interval. Table~\ref{tab:task_overview} summarizes the evaluation setup. Across all experiments, we use a consistent 50\%/50\% split: for each training snapshot, 50\% of observed links are used to form the training network, while the remaining 50\% are held out as positive test links. All reported results are averaged over 10 independent random splits to ensure statistical robustness. Candidate prediction pairs are restricted to physician pairs that are disconnected in the training network. Negative links are sampled at a 20:1 ratio relative to positive links, which exceeds the 5:1 to 10:1 ratios commonly adopted in the link prediction literature~\cite{yang2015evaluating, li2023evaluating}, drawn uniformly at random from physician pairs within two hops that remain unconnected in the evaluation snapshot. To comprehensively assess predictive performance, we consider two complementary tasks: within-period link prediction and cross-period prediction across different temporal windows.

\begin{table}[t]
\centering
\small
\setlength{\tabcolsep}{4pt}
\renewcommand{\arraystretch}{1.12}
\caption{Overview of evaluation tasks for temporal link prediction.}
\label{tab:task_overview}

\begin{tabular}{p{2.2cm} p{2.8cm} p{2.8cm}}
\toprule
Task 
& Training Network 
& Positive Test Links \\
\midrule

Within-period (Task A)
& Same calendar year; 50\% links
& Held-out 50\% from same snapshot \\

\midrule

Cross-period (Task B)
& 30-day short-window network
& New links in 90/180-day window \\

\bottomrule
\end{tabular}
\end{table}

\begin{table*}[t]
\centering
\small
\setlength{\tabcolsep}{2.2pt}
\renewcommand{\arraystretch}{1.05}
\caption{Performance comparison across Tasks A and B under Low/Mid/High average node degree regimes.
For MRP, lower is better ($\downarrow$); for AUPRC and SL@100, higher is better ($\uparrow$).
Best results are shown in \textbf{bold}, and second-best results are \underline{underlined}.}
\label{tab:task_ab_compact}
\begin{tabular}{l ccc ccc ccc ccc ccc ccc}
\toprule
\multirow{2}{*}{Method} &
\multicolumn{9}{c}{Task A} &
\multicolumn{9}{c}{Task B} \\
\cmidrule(lr){2-10}\cmidrule(lr){11-19}
& \multicolumn{3}{c}{AUPRC $\uparrow$} & \multicolumn{3}{c}{MRP $\downarrow$} & \multicolumn{3}{c}{SL@100 $\uparrow$}
& \multicolumn{3}{c}{AUPRC $\uparrow$} & \multicolumn{3}{c}{MRP $\downarrow$} & \multicolumn{3}{c}{SL@100 $\uparrow$} \\
\cmidrule(lr){2-4}\cmidrule(lr){5-7}\cmidrule(lr){8-10}
\cmidrule(lr){11-13}\cmidrule(lr){14-16}\cmidrule(lr){17-19}
& low & mid & high
& low & mid & high
& low & mid & high
& low & mid & high
& low & mid & high
& low & mid & high \\
\midrule

CN
& 0.644 & 0.688 & 0.701
& 0.092 & 0.071 & 0.065
& 4.815 & 5.726 & 6.596
& 0.495 & 0.528 & 0.532
& 0.096 & 0.081 & 0.076
& 5.207 & 6.457 & \underline{7.896} \\

RA
& 0.786 & \underline{0.817} & \underline{0.827}
& 0.080 & 0.058 & \underline{0.051}
& 4.872 & \underline{5.861} & \underline{6.796}
& \underline{0.631} & \underline{0.652} & \underline{0.643}
& \underline{0.084} & \underline{0.068} & \underline{0.063}
& \textbf{5.270} & \underline{6.581} & \textbf{7.993} \\

PA
& 0.384 & 0.366 & 0.332
& 0.142 & 0.142 & 0.153
& 3.914 & 3.982 & 4.028
& 0.225 & 0.235 & 0.221
& 0.198 & 0.184 & 0.165
& 3.891 & 4.158 & 4.366 \\

Jaccard
& 0.091 & 0.147 & 0.208
& 0.379 & 0.267 & 0.207
& 4.057 & 4.557 & 5.070
& 0.113 & 0.152 & 0.172
& 0.340 & 0.268 & 0.231
& 4.274 & 4.872 & 5.265 \\

AA
& 0.710 & 0.739 & 0.746
& 0.091 & 0.069 & 0.062
& 4.822 & 5.746 & 6.627
& 0.542 & 0.564 & 0.562
& 0.097 & 0.080 & 0.075
& 5.207 & 6.459 & 7.870 \\

HP
& 0.080 & 0.095 & 0.118
& 0.301 & 0.248 & 0.207
& 3.496 & 3.321 & 3.565
& 0.095 & 0.105 & 0.105
& 0.254 & 0.232 & 0.228
& 3.973 & 3.973 & 3.634 \\

L3
& \underline{0.805} & 0.792 & 0.780
& \underline{0.058} & \underline{0.051} & \underline{0.051}
& \underline{4.880} & 5.743 & 6.528
& 0.600 & 0.599 & 0.596
& 0.087 & 0.076 & 0.072
& 5.116 & 6.230 & 7.454 \\

LHN
& 0.033 & 0.036 & 0.040
& 0.687 & 0.613 & 0.545
& 2.660 & 2.202 & 2.163
& 0.038 & 0.040 & 0.040
& 0.586 & 0.554 & 0.533
& 3.029 & 2.668 & 2.142 \\

DeepWalk
& 0.463 & 0.517 & 0.531
& 0.119 & 0.101 & 0.093
& 4.102 & 4.982 & 5.742
& 0.302 & 0.348 & 0.311
& 0.143 & 0.118 & 0.110
& 4.522 & 5.121 & 5.503 \\

Node2Vec
& 0.467 & 0.519 & 0.536
& 0.117 & 0.100 & 0.091
& 4.129 & 5.005 & 5.781
& 0.306 & 0.347 & 0.313
& 0.141 & 0.116 & 0.108
& 4.548 & 5.142 & 5.532 \\

GCN
& 0.731 & 0.770 & 0.783
& 0.074 & 0.065 & 0.059
& 4.556 & 5.392 & 6.140
& 0.463 & 0.480 & 0.472
& 0.110 & 0.098 & 0.092
& 4.883 & 5.213 & 5.567 \\

GAT
& 0.461 & 0.560 & 0.640
& 0.125 & 0.095 & 0.084
& 4.090 & 5.061 & 5.944
& 0.291 & 0.356 & 0.330
& 0.148 & 0.120 & 0.110
& 4.413 & 4.982 & 5.302 \\

GIN
& 0.063 & 0.059 & 0.057
& 0.413 & 0.381 & 0.366
& 3.210 & 3.387 & 3.612
& 0.063 & 0.051 & 0.050
& 0.401 & 0.370 & 0.350
& 3.399 & 3.512 & 3.721 \\

GraphSAGE
& 0.538 & 0.652 & 0.727
& 0.104 & 0.084 & 0.071
& 4.244 & 5.214 & 6.019
& 0.384 & 0.443 & 0.401
& 0.122 & 0.102 & 0.093
& 4.701 & 5.234 & 5.612 \\

GraphConv
& 0.182 & 0.191 & 0.168
& 0.216 & 0.199 & 0.181
& 3.846 & 4.042 & 4.271
& 0.084 & 0.068 & 0.065
& 0.265 & 0.240 & 0.230
& 3.813 & 3.962 & 4.081 \\

\midrule
H3 (ours)
& \textbf{0.888} & \textbf{0.890} & \textbf{0.885}
& \textbf{0.047} & \textbf{0.039} & \textbf{0.037}
& \textbf{5.012} & \textbf{5.994} & \textbf{6.999}
& \textbf{0.675} & \textbf{0.680} & \textbf{0.686}
& \textbf{0.071} & \textbf{0.060} & \textbf{0.056}
& \underline{5.237} & \textbf{6.593} & 7.383 \\

\bottomrule
\end{tabular}
\end{table*}

To provide a comprehensive and robust assessment, we adopt a set of complementary ranking-based metrics that capture both global discrimination and top-ranked retrieval quality. Specifically, we report the area under the receiver operating characteristic curve (AUROC) and the area under the precision-recall curve (AUPRC) to evaluate overall ranking performance, with the latter offering a more informative measure when positive links are sparse. To assess source-level retrieval, we additionally report mean reciprocal rank (MRR) and normalized discounted cumulative gain at 100 (NDCG@100) computed over source nodes, which reward systems that rank true positive targets higher within each source's candidate list. We further include mean rank percentile (MRP) as a global measure of relative ranking position across all predictions, source precision at 100 (SP@100) to capture the fraction of relevant links retrieved within the top 100 recommendations per source, and source lift at 100 (SL@100) to quantify the enrichment of true positives relative to a state-level random baseline, averaged first over source nodes within each state and then across states.

We compare H3 against a range of baseline link prediction methods, including common neighbors (CN), resource allocation (RA), preferential attachment (PA), Jaccard index, Adamic--Adar (AA), hub promoted (HP), and Leicht--Holme--Newman (LHN), which are classical local similarity indices~\cite{zhou2009predicting}, as well as L3~\cite{cannistraci2013link}. We further include representation learning and graph neural network approaches such as DeepWalk~\cite{perozzi2014deepwalk}, Node2Vec~\cite{grover2016node2vec}, graph convolutional networks (GCN)~\cite{kipf2016semi}, graph isomorphism networks (GIN)~\cite{xu2018powerful}, graph attention networks (GAT)~\cite{velickovic2017graph}, GraphSAGE~\cite{hamilton2017inductive}, and GraphConv \cite{morris2019weisfeiler}. Among these baselines, only L3 and H3 incorporate link weights into their similarity computations; all other methods operate solely on the binary adjacency structure and do not utilize link weights. For GNN-based methods (GCN, GIN, GAT, GraphSAGE, and GraphConv), node features are initialized using Node2Vec embeddings trained on the training network.

\subsection{Overall Performance across Tasks}
\label{sec:overall_performance}

Across Task A and Task B, link prediction performance is evaluated using global discrimination metrics (AUROC, AUPRC), source-level retrieval quality (Source MRR, Source NDCG@100), and ranking-based metrics (NDCG@100, Mean Rank Percentile, SP@100, SL@100) as summarized in Fig. \ref{fig:link_prediction_heatmap}. The results reveal three clear trends. First, H3 consistently outperforms all competing methods, with particularly strong gains in early-retrieval metrics (Source MRR, NDCG@100, SL@100), which are most relevant for recommending high-confidence referrals. Second, traditional local heuristics such as CN, AA, RA, and L3 maintain robust performance, often surpassing embedding- and GNN-based approaches, highlighting the value of direct structural cues in the sparse, hub-dominated referral network. Third, representation learning and GNN-based methods, including DeepWalk, Node2Vec, GCN, GAT, GIN, GraphSAGE, and GraphConv, tend to underperform, especially under class imbalance, indicating that latent embeddings and extensive message passing may not generalize well to the specific patterns of physician referrals. Overall, these observations position H3 as the most reliable method for ranking potential referral links and motivate the more detailed breakdowns presented in the following subsections.


\subsection{Adaptability to Sparsity}
\label{sec:adaptability_sparsity}
Physician referral networks are intrinsically sparse and long-tailed. Many physicians, especially in small states, have very limited observed interactions, creating a severe imbalance between non-links and future links. This regime is notoriously difficult for parameterized link predictors, as scarcity and skewed distributions amplify estimation variance and lead to unstable ranking behavior~\cite{ma2025class}. Evaluating robustness under such conditions is not merely a stress test but a clinically relevant requirement, since sparse regions correspond to realistic deployment settings where the system must still produce actionable recommendations from limited evidence.

\begin{table*}[t]
\centering
\small
\setlength{\tabcolsep}{4pt}
\renewcommand{\arraystretch}{1.15}
\caption{Hyperparameter sensitivity and ablation configurations for H3.
The default setting is used as the baseline unless otherwise specified.}
\label{tab:h3_sensitivity_config}
\begin{tabular}{clll}
\toprule
Group & Parameter(s) & Tested Values & Rationale \\
\midrule

G1 &
\begin{tabular}[c]{@{}l@{}}Degree Normalization\\(\textit{Normalization Sensitivity})\end{tabular} &
\begin{tabular}[c]{@{}l@{}}
No Norm: $(0.0)$ \\
Weak: $(0.2)$ \\
Sqrt: $(0.5)$ \\
Default: $(0.8 / 0.2)$ \\
Strong: $(1.0)$
\end{tabular} &
\begin{tabular}[c]{@{}l@{}}
Evaluates the necessity and strength of suppressing \\
high-degree (hub) nodes. Tests whether hub \\
dominance harms referral link prediction.
\end{tabular} \\

\addlinespace[4pt]

G2 &
\begin{tabular}[c]{@{}l@{}}Path Weight Impact\\(\textit{Weight Ablation})\end{tabular} &
\begin{tabular}[c]{@{}l@{}}
Unweighted: $0.0$ \\
Dampened: $0.3$ (Default) \\
Linear: $1.0$ \\
Amplified: $2.0$
\end{tabular} &
\begin{tabular}[c]{@{}l@{}}
Validates whether referral intensity (link weight) \\
provides additional predictive power beyond \\
pure topology. $0.0$ corresponds to a strict ablation.
\end{tabular} \\

\addlinespace[4pt]

G3 &
\begin{tabular}[c]{@{}l@{}}Directionality\\(\textit{Asymmetry Analysis})\end{tabular} &
\begin{tabular}[c]{@{}l@{}}
Pure Reverse: $0.0$ \\
Reverse-biased: $0.2$ \\
Balanced: $0.5$ \\
Forward-biased: $0.8$ \\
Pure Forward: $1.0$
\end{tabular} &
\begin{tabular}[c]{@{}l@{}}
Examines whether referral relationships exhibit \\
directional asymmetry in healthcare networks, \\
and whether forward flow dominates prediction.
\end{tabular} \\

\addlinespace[4pt]

G4 &
\begin{tabular}[c]{@{}l@{}}Penalty Term\\(\textit{Structural Ablation})\end{tabular} &
\begin{tabular}[c]{@{}l@{}}
Off: $0.0$ \\
Default: $0.5$ \\
Strong: $1.0$
\end{tabular} &
\begin{tabular}[c]{@{}l@{}}
Ablates the proposed redundancy penalty term $p_m^{(y)}$. \\
If performance drops at $0.0$, the penalty is beneficial; \\
otherwise it may be redundant.
\end{tabular} \\
\bottomrule
\end{tabular}
\end{table*}

To quantify the effect of network connectivity, we partition states into low, mid, and high categories based on tertiles of the average node degree in the training network, which captures connectivity independently of network size. Table~\ref{tab:task_ab_compact} summarizes the adaptability results for H3 and all baseline methods. Across Tasks~A and~B, learning-based methods exhibit pronounced degradation or volatility in low-degree states, particularly on AUPRC and SL@100. This aligns with findings that tail node pairs, which are both frequent and systematically harder for GNN-style predictors, dominate long-tailed link prediction~\cite{wang2024optimizing}. In contrast, H3 achieves the highest AUPRC in low-degree states for both Task~A (0.888) and Task~B (0.675). On SL@100, H3 leads all methods in the low-degree regime for Task~A (5.012) and remains highly competitive in Task~B low-degree states (5.237), where RA attains the top score (5.270). H3 also attains the best MRP in most low- and mid-degree conditions, with GraphConv recording the lowest MRP in a subset of regimes owing to its conservative, degree-smoothed scoring that suppresses false positives in small networks at the cost of discriminative power. The consistent AUPRC advantage of H3, together with its strong SL@100 performance, indicates superior global and top-list ranking quality in sparse settings, attributable to its near parameter-free design and reliance on explicit three-hop structural motifs, which provide stable evidence when two-hop statistics or learned embeddings are noisy or under-determined.

Moreover, in mid- and high-degree states, most methods improve monotonically as network connectivity increases. Several learning-based baselines, most notably GCN, which reaches AUPRC 0.783 (Task~A, high) and 0.472 (Task~B, high), close part of the gap with H3 as more training signal becomes available, and GraphConv achieves competitive or leading MRP across several mid- and high-degree regimes for both tasks. Nonetheless, H3 retains the best AUPRC across all three regimes in both tasks and the best SL@100 across all three regimes in Task~A, while remaining consistently competitive on SL@100 in Task~B despite being outperformed there by RA. These results demonstrate that H3 maintains stable global ranking performance while preserving its robustness advantage in sparse settings.

\subsection{Hyperparameter Sensitivity}
\label{sec:hparam_sensitivity}
To evaluate the robustness of H3 and disentangle the contributions of its design components, we conduct a structured hyperparameter sensitivity analysis over 17 interpretable configurations. Rather than performing an exhaustive sweep across all states, we select a representative stratified subset. States are partitioned into small, medium, and large groups based on tertiles of physician count in the training network, and roughly one third of states are sampled from each group. This approach preserves heterogeneity in network scale, degree skewness, and sparsity while keeping the evaluation computationally tractable. All configurations follow the same evaluation protocol as the main experiments.

\begin{figure*}[t]
\centering
\includegraphics[width=0.9\textwidth]{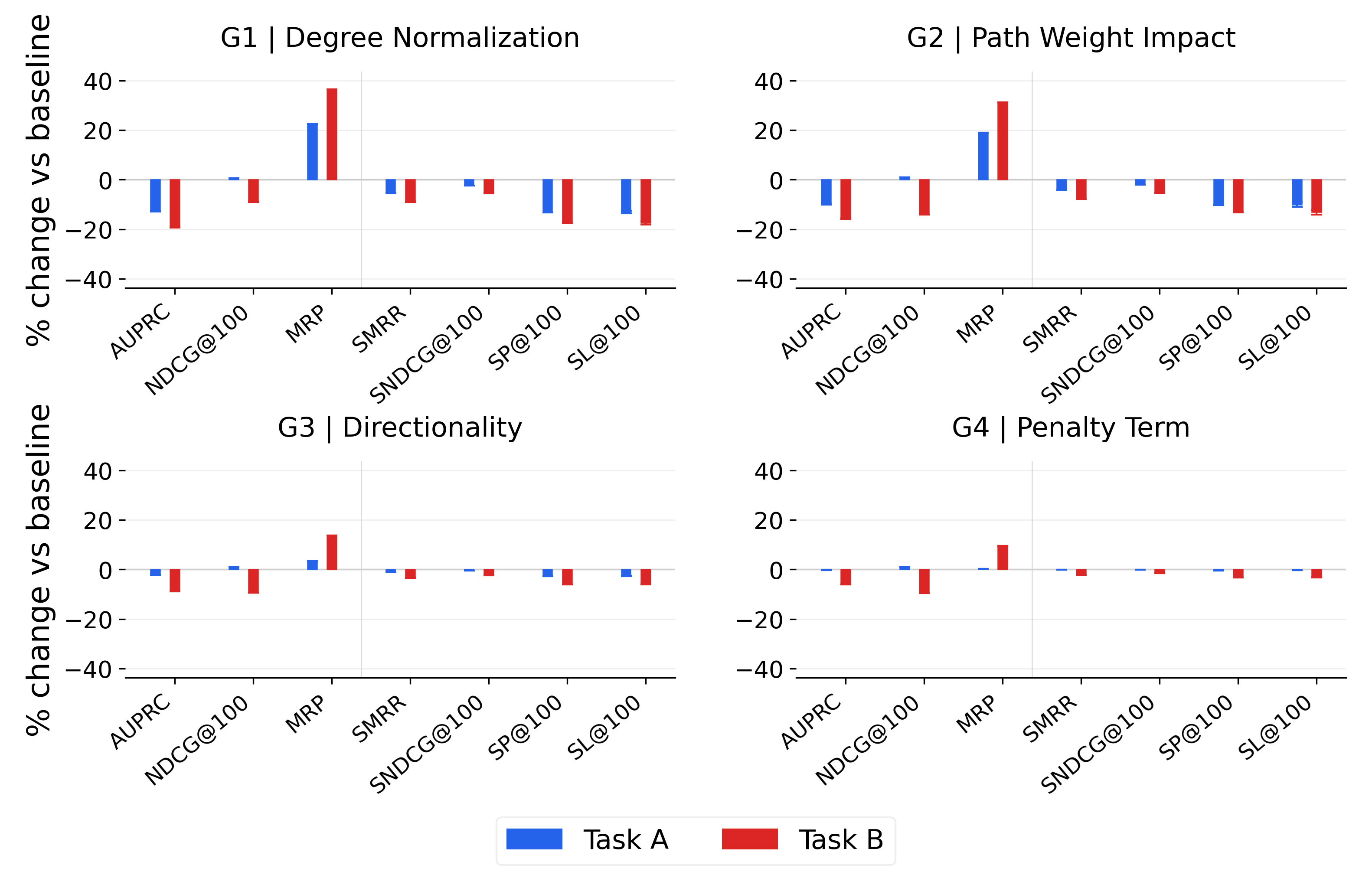}
\caption{Hyperparameter sensitivity of H3 grouped by design component.
Each panel (G1--G4) corresponds to a parameter family defined in Table~\ref{tab:h3_sensitivity_config}.}
\label{fig:hparam_sensitivity}
\end{figure*}

Table~\ref{tab:h3_sensitivity_config} summarizes the explored hyperparameter space, organized into four groups according to their structural roles. Group~G1 (degree normalization) controls hub suppression strength via $(\beta, \gamma)$. Group~G2 (path weight impact) varies the path exponent $\alpha$, determining how strongly high-weight three-hop paths dominate aggregation. Group~G3 (directionality) adjusts the forward-reverse mixing coefficient $\epsilon$, probing whether referral prediction is inherently asymmetric. Group~G4 (penalty term) ablates the redundancy penalty $\eta$ applied to intermediate nodes. Each group corresponds to a distinct modeling hypothesis rather than an arbitrary tuning dimension. Fig.~\ref{fig:hparam_sensitivity} reveals several consistent patterns. Path weight modeling (G2) is the most performance-critical component. Varying $\alpha$ induces the largest spread across all metrics and both tasks, indicating that how H3 integrates referral intensity along three-hop paths fundamentally determines predictive accuracy. Degree normalization (G1) is essential but not brittle: removing hub suppression leads to clear degradation, yet a broad plateau of near-optimal settings exists, suggesting that H3 does not rely on finely tuned penalties to control hub effects. Directionality (G3) exhibits asymmetric but stable gains: forward-biased settings consistently outperform symmetric or reverse-only variants, confirming that referral dynamics encode directional flow, while limited variance indicates robustness to moderate mis-specification of $\alpha$. Finally, the redundancy penalty (G4) provides consistent but moderate improvements: disabling the penalty ($\eta=0$) leads to systematic drops, supporting the hypothesis that suppressing redundant intermediate connectors stabilizes ranking under sparse and noisy conditions.

\subsection{Robustness to Physician Mobility}
\label{sec:robustness_mobility}

Expanding the observation window from a short snapshot to a longer period inherently increases incidental patient mobility and hub mixing, amplifying parallel multi-hop connectivity that does not reflect stable clinical coordination. To quantify this effect, we analyze Task~B (cross-period prediction) and define the link expansion ratio $r = |E_{\text{long}}| / |E_{\text{short}}|$ for each state, where $|E_{\text{short}}|$ and $|E_{\text{long}}|$ denote the number of weighted links in the short- and long-window networks, respectively. A higher $r$ indicates a greater proportion of low-weight, hub-mediated links that may reflect incidental patient mobility rather than persistent referral relationships.

\begin{figure}[t]
    \centering
    \includegraphics[width=\columnwidth]{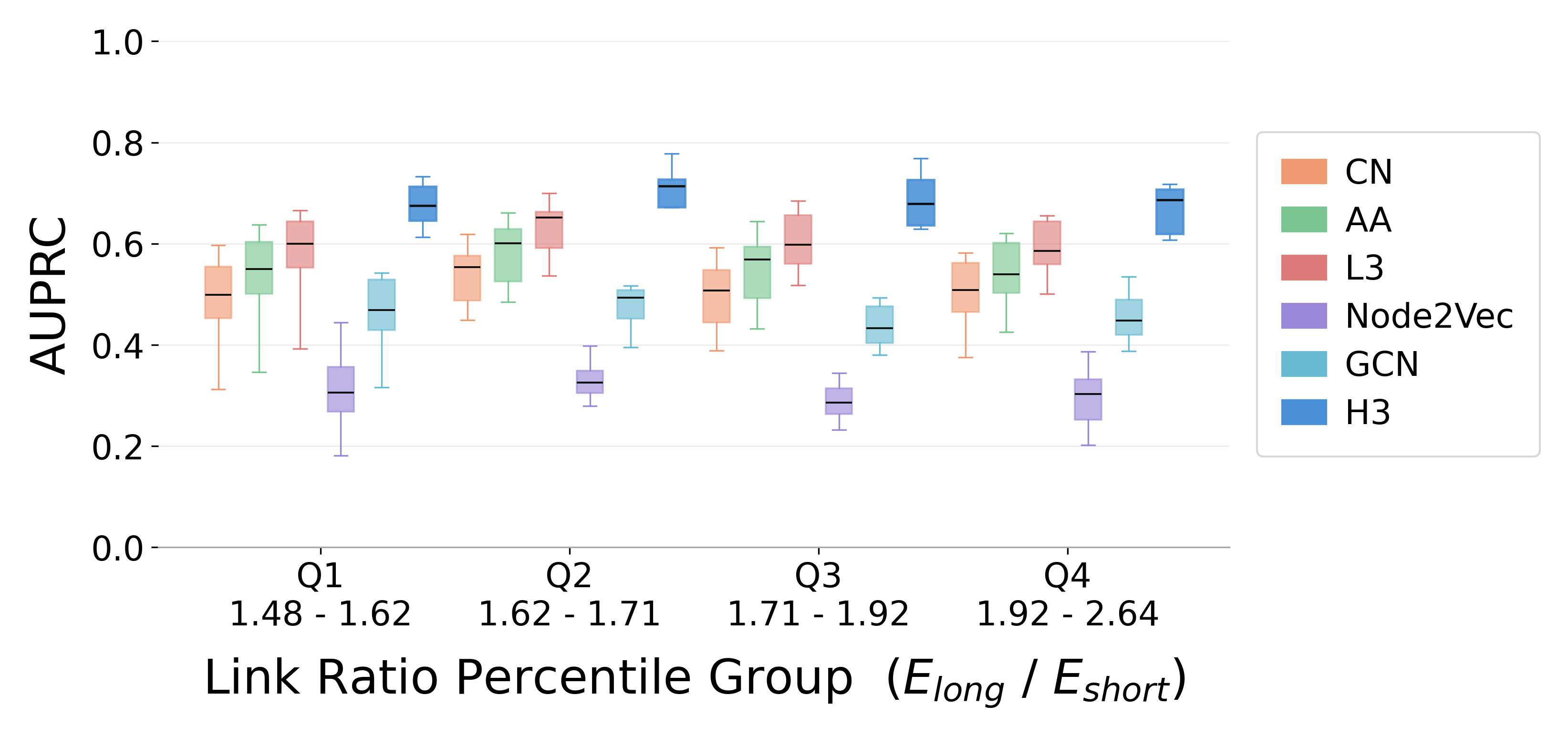}
    \caption{AUPRC across link expansion ratio quartiles $r$ under Task B. Each box shows the 5--95 percentile range over states.}
\label{fig:redundancy_precision}
\end{figure}

Fig.~\ref{fig:redundancy_precision} plots  AUPRC across link expansion ratio quartiles for representative methods under Task B. CN and AA exhibit moderate but consistent decline as $r$ increases, reflecting their sensitivity to hub-mediated shortcut routes introduced by longer observation windows. Node2Vec shows pronounced degradation across quartiles, suggesting that embedding-based methods are particularly vulnerable to distributional shift when incidental mobility inflates the edge set. L3 shows intermediate sensitivity: its degree normalization partially mitigates hub inflation, but the absence of an explicit redundancy penalty leaves it vulnerable at high $r$ values. GCN improves slightly with higher $r$, likely benefiting from the denser training signal, yet its median AUPRC remains substantially below H3 across all quartiles. In contrast, H3 maintains consistently high AUPRC with narrow interquartile ranges across all four quartiles, demonstrating stable ranking fidelity as the network densifies. This robustness is attributable to the redundancy penalty $p_k^{(j)}$, which downweights repeated intermediate evidence that would otherwise dominate in dense hub regions.

\section{Discussion}\label{sec:dis}
\subsection{Micro-Level Case Study}
\label{sec:case_study}
\begin{table*}[t]
\centering
\small
\setlength{\tabcolsep}{4pt}
\renewcommand{\arraystretch}{1.15}
\caption{Micro-level Structural Analysis for Case Study}
\label{tab:integrated_case_study}
\begin{tabular}{cc c c c cc l}
\toprule
\multicolumn{2}{c}{Target Pair} & 
H3 & 
L3 &
True & 
CN & 
\multicolumn{1}{c}{Top Common Neighbors} & 
3-Hop Path \\
Source ($u$) & Target ($v$) & 
Rank & 
Rank &
Label & 
Count & 
\multicolumn{1}{c}{(Mega-Hubs)} & 
Count \\
\midrule

\multicolumn{8}{c}{\textit{Case 1: Successful Noise Filtration (H3 identifies hidden intent despite hubs)}} \\
\midrule
$S_1$ & $S_2$ & \textbf{11} & $\ge 1000$ & 1 & 52 & $H_1$ (1121), $H_2$ (617), $H_3$ (600) & 10,199 \\
$S_3$ & $S_4$ & \textbf{14} & $\ge 1000$ & 1 & 57 & $H_1$ (1121), $H_2$ (617), $H_3$ (600) & 9,453 \\
$S_5$ & $S_6$ & \textbf{16} & $\ge 1000$ & 1 & 49 & $H_1$ (1121), $H_3$ (600), $H_4$ (468) & 9,024 \\
$S_7$ & $S_8$ & \textbf{7} & $\ge 1000$ & 1 & 49 & $H_1$ (1121), $H_2$ (617), $H_4$ (468) & 8,765 \\
$S_9$ & $S_{10}$ & \textbf{4} & $\ge 1000$ & 1 & 52 & $H_1$ (1121), $H_2$ (617), $H_3$ (600) & 8,420 \\

\midrule
\multicolumn{8}{c}{\textit{Case 2: Signal Over-Amplification (False Positives: High Rank, No Referral)}} \\
\midrule
$F_{P1}$ & $F_{P2}$ & 3 & 21 & 0 & 103 & $H_1$ (1121), $H_2$ (617), $H_3$ (600) & 11,991 \\
$F_{P3}$ & $F_{P4}$ & 40 & 72 & 0 & 71 & $H_1$ (1121), $H_2$ (617), $H_3$ (600) & 8,204 \\
$F_{P5}$ & $F_{P6}$ & 49 & 21 & 0 & 37 & $H_1$ (1121), $H_3$ (600), $H_5$ (455) & 8,072 \\
$F_{P7}$ & $F_{P8}$ & 54 & 46 & 0 & 55 & $H_1$ (1121), $H_2$ (617), $H_3$ (600) & 8,542 \\
$F_{P9}$ & $F_{P10}$ & 77 & 113 & 0 & 61 & $H_1$ (1121), $H_2$ (617), $H_3$ (600) & 7,788 \\

\midrule
\multicolumn{8}{c}{\textit{Case 3: Signal Over-Penalization (False Negatives: Low Rank, True Referral)}} \\
\midrule
$F_{N1}$ & $F_{N2}$ & $\ge 1000$ & $\ge 1000$ & 1 & 94 & $H_1$ (1121), $H_2$ (617), $H_3$ (600) & 9,314 \\
$F_{N3}$ & $F_{N4}$ & $\ge 1000$ & $\ge 1000$ & 1 & 75 & $H_1$ (1121), $H_2$ (617), $H_3$ (600) & 11,084 \\
$F_{N5}$ & $F_{N6}$ & $\ge 1000$ & $\ge 1000$ & 1 & 72 & $H_1$ (1121), $H_2$ (617), $H_3$ (600) & 9,984 \\
$F_{N7}$ & $F_{N8}$ & $\ge 1000$ & $\ge 1000$ & 1 & 110 & $H_1$ (1121), $H_2$ (617), $H_3$ (600) & 11,364 \\
$F_{N9}$ & $F_{N10}$ & $\ge 1000$ & $\ge 1000$ & 1 & 64 & $H_1$ (1121), $H_2$ (617), $H_3$ (600) & 10,922 \\

\bottomrule
\end{tabular}
\end{table*}
To further understand the predictive behavior of H3, we conduct a micro-level structural analysis as a case study,  highlighting both the strengths and structural boundaries of topology-based referral prediction (Table~\ref{tab:integrated_case_study}). We examine three categories of node pairs: true positives recovered by H3 and L3, false positives ranked highly by both methods, and false negatives missed by both. These true-positive pairs share around 50 common neighbors but are embedded in hub-dominated neighborhoods (e.g., $H_1$ with degree 1121). L3 aggregates thousands of hub-mediated three-hop paths without redundancy control, assigning ranks beyond 1000 to all five pairs. H3 introduces the redundancy penalty $p_k^{(j)}$, which discounts intermediaries that connect promiscuously to the same target, allowing these pairs to be recovered within the top-50. Several false-positive pairs receive relatively high ranks under L3 because unweighted path aggregation treats all hub-mediated paths equally. H3 partially mitigates this effect by incorporating link-weight-sensitive path aggregation, though structural similarity between true and spurious pairs still creates ambiguity. Both H3 and L3 rank all false-negative pairs beyond 1000. These pairs share nearly identical macro-structural profiles with false positives, featuring 8,000--12,000 three-hop paths routed through the same mega-hubs. This indicates a fundamental limitation of purely topological prediction.

These results  establish that H3 robustly captures referral patterns in sparse and moderately hub-noisy environments where L3 fails, but that both methods encounter an information bottleneck at the extremes of the degree distribution. This boundary is consistent with theoretical findings that every topological feature possesses an inherent prediction performance upper bound determined by the network's structural resolution~\cite{ran2024maximum}, and that high-degree hub nodes introduce confounding biases that decouple topological proximity from true relational semantics~\cite{sosa2024elucidating}. Overcoming this boundary would require non-topological signals such as clinical specialty alignment, geographic proximity, or institutional affiliation, which fall outside the scope of claims-derived network data and represent a natural direction for future work~\cite{grover2016node2vec}.

\subsection{Generalizability to Other Domains}
\label{sec:generalization}

To assess whether the advantage of H3 is specific to physician referral networks or reflects a more general structural principle, we further evaluate H3 on two representative benchmarks from the Open Graph Benchmark (OGB)~\cite{hu2020ogb}: \texttt{ogbl-collab} (academic collaboration) and \texttt{ogbl-citation2} (paper citation). These datasets differ from healthcare networks in semantics, scale, and supervision density, providing a meaningful test of cross-domain transferability. Full results are reported in Table \ref{tab:ogb_results} in the appendix. Despite having zero trainable parameters, H3 consistently outperforms classical proximity indices (CN, AA, Jaccard) and achieves competitive performance relative to embedding-based methods such as Node2Vec on both benchmarks. Notably, H3 improves upon L3 across both datasets, achieving 66.86\% Hits@50 on \texttt{ogbl-collab} and 54.83\% MRR on \texttt{ogbl-citation2}, a benchmark where learned models typically hold a considerable advantage. These results do not approach the state-of-the-art set by large-scale learned models on OGB leaderboards. However, they show that a parameter-free heuristic can meaningfully narrow the gap relative to classical baselines, suggesting that the three-hop inductive bias with redundancy control captures structural regularities that extend beyond healthcare networks.

Beyond social and information networks, similar structural principles emerge in biological domains. Large-scale evaluations of protein-protein interaction (PPI) networks~\cite{wang2023assessment} show that across diverse organisms, including \textit{A. thaliana}, \textit{C. elegans}, yeast, and the human HuRI network, three-hop path–based predictors consistently rank among the strongest non-parametric methods, often matching or outperforming more complex models (Table \ref{tab:ppi_benchmark_h3} in the appendix). The concordance between these PPI results and the OGB findings reported here suggests that H3 inherits a broadly transferable inductive bias. It emphasizes mid-range structural flow patterns while suppressing spurious hub-dominated shortcuts. By explicitly modeling three-hop connectivity with redundancy control, H3 captures a form of mesoscopic organization, consisting of indirect, degree-regulated paths through intermediaries, that recurs across social, biological, and information networks, enabling robust generalization without task-specific training or large parameter budgets.

\subsection{Limitations and Future Work}
\label{sec:limitation_future_work}

Although H3 demonstrates robust performance and interpretability across diverse referral networks, several limitations merit discussion. As a purely topological heuristic, H3 is inherently blind to node-level semantics. Our micro-level case study reveals that in the mega-hub regime, structurally indistinguishable topologies can produce divergent clinical outcomes, because H3 cannot differentiate whether an extensive three-hop pathway connects two redundant specialists or a natural PCP-specialist referral pair. Additionally, H3 strictly relies on continuous paths up to three hops, making it subject to the cold-start problem shared by all proximity-based predictors. For newly practicing physicians or isolated rural providers, historical multi-hop patient flows may be entirely absent. Bridging this semantic gap and addressing zero-path scenarios will require integrating H3 scores as structural features within attribute-aware frameworks (e.g., gradient-boosted trees or lightweight GNN architectures) that can jointly leverage topology and physician metadata such as specialty\cite{xiong2026survey}, geography, and institutional affiliation, while falling back on content-based signals when topological evidence is unavailable~\cite{zheng2025survey}. Beyond methodological extensions, our current framework models temporal dynamics through discrete, pre-defined snapshots, whereas healthcare delivery is a continuous process with shifting seasonal demands and evolving hospital capacities~\cite{ochoa2022graph}. Future work could explore continuous-time dynamic network formulations that incorporate temporal decay functions over path weights to more accurately capture the transient nature of patient-sharing relationships~\cite{paul2024systematic}. In summary, while H3 establishes a simple yet powerful structural baseline tailored for the disassortative topology of healthcare networks, evolving it into a hybrid, attribute-aware, and continuous-time learning system represents a promising direction  for advancing next-generation clinical decision-support research.

\section{Conclusion}\label{sec:conclusion}

In this article, we introduce H3, a weighted three-hop scoring function designed to address the fundamental challenges of link prediction in sparse, hub-dominated physician referral networks. Extensive evaluation on nationwide Medicare data across all U.S. states demonstrates that H3 consistently outperforms traditional heuristics and state-of-the-art machine learning methods across the majority of evaluation settings, achieving superior performance on early retrieval metrics that are critical for high-confidence clinical recommendations. Structural analyses reveal that H3's effectiveness stems from a principled inductive bias aligned with the disassortative, sparse topology of healthcare delivery networks, where empirical evidence from network science supports the effectiveness of three-hop indices over conventional two-hop alternatives. While H3 is inherently limited to topological signals and cannot resolve ambiguity in extreme hub-dominated regimes, these boundaries delineate clear opportunities for hybrid approaches that combine structural path evidence with clinical metadata.

\section*{References}
\bibliographystyle{IEEEtran}
\bibliography{example_bib}

\appendices


\begin{table*}[th]
\centering
\small
\setlength{\tabcolsep}{4.5pt}
\renewcommand{\arraystretch}{1.1}
\caption{%
  Performance comparison on OGB link prediction benchmarks.
  \textbf{C} = \texttt{ogbl-collab} (academic collaboration);
  \textbf{Ci} = \texttt{ogbl-citation2} (paper citation).
  \textbf{\#Params (C / Ci)}: number of trainable parameters on each dataset
  \textbf{Hits@50}: percentage of true links ranked within the top-50.
  \textbf{MRR}: Mean Reciprocal Rank.%
}
\label{tab:ogb_results}
\begin{tabular}{l c c c}
\toprule
\textbf{Method}
  & \textbf{\#Params (C / Ci)}
  & \textbf{Hits@50 (C)}
  & \textbf{MRR (Ci)} \\
\midrule
\multicolumn{4}{l}{\emph{State-of-the-art learning-based models}} \\
\midrule
HyperFusion        & 1.06B / --      & 71.29 & --    \\
GIDN@YITU          & 60.45M / --     & 70.96 & --    \\
GraphGPT (d1n30)   & -- / 133.10M    & --    & 93.05 \\
MPLP               & -- / 749.76M    & --    & 90.72 \\
SIEG               & -- / 2.43M      & --    & 90.18 \\
\midrule
\multicolumn{4}{l}{\emph{Representative GNN and embedding baselines}} \\
\midrule
SEAL        & 0.50M / 0.26M    & 64.74 & 87.67 \\
GraphSAGE   & 0.46M / 0.46M    & 48.10 & 82.60 \\
GCN         & 0.30M / 0.30M    & 44.75 & 84.74 \\
Node2Vec    & 30.32M / 374.91M & 48.88 & 61.41 \\
DeepWalk    & 61.39M / 61.39M  & 50.37 & 60.42 \\
\midrule
\multicolumn{4}{l}{\emph{Zero-parameter methods}} \\
\midrule
Common Neighbors (CN) & 0 / 0       & 61.37 & 51.47 \\
Adamic--Adar (AA)     & 0 / 0       & 64.17 & 51.89 \\
Jaccard Index         & 0 / 0       & 50.50 & 50.98 \\
Matrix Factorization  & 0 / 281.11M & 38.86 & 51.86 \\
\midrule
\textbf{H3 (Ours)}
  & \textbf{0 / 0}
  & \textbf{66.86}
  & \textbf{54.83} \\
\bottomrule
\end{tabular}
\end{table*}

\begin{table*}[th]
\centering
\small
\setlength{\tabcolsep}{3.5pt}
\renewcommand{\arraystretch}{1.08}
\caption{%
  PPI link prediction benchmark results (10-fold CV) across four biological
  networks.Metrics follow:
  \textbf{P@500} = Precision at the top-500 predicted links;
  \textbf{NDCG} = Normalized Discounted Cumulative Gain.%
}
\label{tab:ppi_benchmark_h3}
\resizebox{\textwidth}{!}{%
\begin{tabular}{lcccccccccccccccc}
\toprule
\multirow{2}{*}{Method} &
\multicolumn{4}{c}{\textit{A. thaliana}} &
\multicolumn{4}{c}{\textit{C. elegans}} &
\multicolumn{4}{c}{\textit{S. cerevisiae} (Yeast)} &
\multicolumn{4}{c}{\textit{H. sapiens} (HuRI)} \\
\cmidrule(lr){2-5}\cmidrule(lr){6-9}\cmidrule(lr){10-13}\cmidrule(lr){14-17}
 & AUROC & AUPRC & P@500 & NDCG
 & AUROC & AUPRC & P@500 & NDCG
 & AUROC & AUPRC & P@500 & NDCG
 & AUROC & AUPRC & P@500 & NDCG \\
\midrule

\multicolumn{17}{l}{\textbf{Heuristic Methods}}\\
CN       & 0.60 & 0.00 & 0.00 & 0.44 & 0.56 & 0.00 & 0.01 & 0.43 & 0.60 & 0.00 & 0.04 & 0.45 & 0.76 & 0.00 & 0.00 & 0.56 \\
RA       & 0.60 & 0.00 & 0.00 & 0.44 & 0.56 & 0.00 & 0.00 & 0.42 & 0.60 & 0.00 & 0.04 & 0.42 & 0.76 & 0.00 & 0.00 & 0.55 \\
PA       & 0.75 & 0.00 & 0.00 & 0.42 & 0.59 & 0.00 & 0.00 & 0.36 & 0.59 & 0.00 & 0.04 & 0.42 & 0.89 & 0.00 & 0.00 & 0.53 \\
JC       & 0.61 & 0.00 & 0.00 & 0.45 & 0.56 & 0.00 & 0.00 & 0.44 & 0.61 & 0.00 & 0.04 & 0.45 & 0.77 & 0.00 & 0.00 & 0.54 \\
AA       & 0.60 & 0.00 & 0.00 & 0.44 & 0.56 & 0.00 & 0.00 & 0.42 & 0.60 & 0.00 & 0.04 & 0.42 & 0.76 & 0.00 & 0.00 & 0.55 \\
Katz     & 0.76 & 0.00 & 0.00 & 0.40 & 0.59 & 0.00 & 0.00 & 0.35 & 0.59 & 0.00 & 0.04 & 0.40 & 0.90 & 0.00 & 0.00 & 0.53 \\
SIM      & 0.85 & 0.00 & 0.01 & 0.42 & 0.80 & 0.00 & 0.00 & 0.37 & 0.85 & 0.00 & 0.05 & 0.42 & 0.89 & 0.00 & 0.00 & 0.50 \\
Ensemble & 0.84 & 0.00 & 0.00 & 0.42 & 0.77 & 0.00 & 0.00 & 0.37 & 0.84 & 0.00 & 0.05 & 0.42 & 0.91 & 0.00 & 0.00 & 0.53 \\
\addlinespace[2pt]

\multicolumn{17}{l}{\textbf{Similarity-based Methods}}\\
MPS(T)   & 0.77 & 0.02 & 0.09 & 0.52 & 0.59 & 0.00 & 0.01 & 0.39 & 0.59 & 0.00 & 0.05 & 0.39 & 0.91 & 0.06 & 0.17 & 0.65 \\
MPS(R1)  & 0.77 & 0.02 & 0.10 & 0.52 & 0.59 & 0.00 & 0.02 & 0.40 & 0.59 & 0.01 & 0.05 & 0.42 & 0.92 & 0.04 & 0.30 & 0.64 \\
RNM      & 0.85 & 0.04 & 0.12 & 0.57 & 0.71 & 0.02 & 0.05 & 0.45 & 0.71 & 0.02 & 0.05 & 0.45 & 0.94 & 0.06 & 0.34 & 0.68 \\
L3       & 0.88 & 0.04 & 0.11 & 0.58 & 0.76 & 0.02 & 0.05 & 0.45 & 0.76 & 0.02 & 0.06 & 0.45 & 0.95 & 0.04 & 0.25 & 0.67 \\
L$^{*}$3(t) & 0.53 & 0.00 & 0.04 & 0.40 & 0.52 & 0.00 & 0.02 & 0.35 & 0.52 & 0.01 & 0.03 & 0.36 & 0.89 & 0.01 & 0.13 & 0.56 \\
\addlinespace[2pt]

\multicolumn{17}{l}{\textbf{Matrix Factorization}}\\
SBM      & 0.88 & 0.00 & 0.00 & 0.43 & 0.81 & 0.00 & 0.01 & 0.42 & 0.81 & 0.00 & 0.01 & 0.42 & 0.93 & 0.00 & 0.01 & 0.58 \\
RepGSP1  & 0.39 & 0.00 & 0.03 & 0.40 & 0.46 & 0.00 & 0.01 & 0.38 & 0.46 & 0.00 & 0.00 & 0.40 & 0.34 & 0.00 & 0.01 & 0.48 \\
RepGSP2  & 0.80 & 0.03 & 0.10 & 0.56 & 0.66 & 0.01 & 0.04 & 0.47 & 0.66 & 0.01 & 0.05 & 0.47 & 0.92 & 0.04 & 0.26 & 0.65 \\
NNMF     & 0.58 & 0.00 & 0.11 & 0.35 & 0.55 & 0.00 & 0.31 & 0.31 & 0.55 & 0.00 & 0.33 & 0.32 & 0.69 & 0.00 & 0.06 & 0.45 \\
GLEE     & 0.51 & 0.00 & 0.00 & 0.38 & 0.51 & 0.00 & 0.00 & 0.37 & 0.51 & 0.00 & 0.00 & 0.39 & 0.50 & 0.00 & 0.00 & 0.44 \\
GRLC     & 0.66 & 0.00 & 0.00 & 0.52 & 0.49 & 0.00 & 0.10 & 0.70 & 0.49 & 0.00 & 0.24 & 0.58 & 0.50 & 0.00 & 0.00 & 0.43 \\
\addlinespace[2pt]

\multicolumn{17}{l}{\textbf{Deep Learning Methods}}\\
cGAN1    & 0.64 & 0.02 & 0.01 & 0.61 & 0.69 & 0.01 & 0.01 & 0.58 & 0.69 & 0.02 & 0.02 & 0.57 & 0.71 & 0.03 & 0.00 & 0.71 \\
cGAN2    & 0.80 & 0.01 & 0.02 & 0.47 & 0.78 & 0.00 & 0.01 & 0.38 & 0.78 & 0.01 & 0.03 & 0.43 & 0.73 & 0.01 & 0.02 & 0.57 \\
SkipGNN  & 0.82 & 0.00 & 0.00 & 0.46 & 0.72 & 0.00 & 0.00 & 0.41 & 0.72 & 0.00 & 0.00 & 0.41 & 0.87 & 0.00 & 0.00 & 0.56 \\
SEAL     & 0.92 & 0.01 & 0.04 & 0.52 & 0.88 & 0.01 & 0.02 & 0.44 & 0.88 & 0.01 & 0.03 & 0.44 & 0.94 & 0.01 & 0.01 & 0.59 \\
\addlinespace[2pt]

\multicolumn{17}{l}{\textbf{Random Walk Methods}}\\
ACT      & 0.83 & 0.00 & 0.00 & 0.40 & 0.82 & 0.00 & 0.00 & 0.37 & 0.82 & 0.00 & 0.00 & 0.37 & 0.90 & 0.00 & 0.00 & 0.50 \\
RWR      & 0.78 & 0.00 & 0.00 & 0.43 & 0.60 & 0.00 & 0.00 & 0.35 & 0.60 & 0.00 & 0.00 & 0.35 & 0.91 & 0.00 & 0.00 & 0.53 \\
SimRank  & 0.67 & 0.00 & 0.00 & 0.36 & 0.55 & 0.00 & 0.00 & 0.33 & 0.55 & 0.00 & 0.00 & 0.33 & 0.82 & 0.00 & 0.00 & 0.47 \\
DNN+node2vec & 0.88 & 0.00 & 0.00 & 0.43 & 0.71 & 0.00 & 0.00 & 0.37 & 0.71 & 0.00 & 0.00 & 0.37 & 0.90 & 0.00 & 0.00 & 0.54 \\
RW       & 0.73 & 0.00 & 0.00 & 0.42 & 0.59 & 0.00 & 0.00 & 0.38 & 0.59 & 0.00 & 0.00 & 0.38 & 0.85 & 0.00 & 0.01 & 0.52 \\
\midrule

{\textbf{H3 (Ours)}} &
\textbf{0.89} & \textbf{0.04} & \textbf{0.13} & \textbf{0.58} &
\textbf{0.76} & \textbf{0.02} & \textbf{0.04} & \textbf{0.46} &
\textbf{0.77} & \textbf{0.02} & \textbf{0.05} & \textbf{0.45} &
\textbf{0.95} & \textbf{0.05} & \textbf{0.23} & \textbf{0.67} \\
\bottomrule
\end{tabular}%
}
\end{table*}

\end{document}